\documentclass[preprint,5p,times,twocolumn]{elsarticle}
\usepackage{graphics}
\usepackage{subfigure}
\usepackage{amssymb}
\usepackage{amsmath}
\usepackage[dvipsnames]{xcolor}
\usepackage[export]{adjustbox}
\usepackage{subcaption}
\usepackage{graphicx}
\usepackage{booktabs}
\usepackage{multicol, multirow}
\usepackage{tcolorbox}

\begin{document}
\begin{frontmatter}
\title{Phase transition and universality of the  majority-rule model on complex networks}

\author[brin,uty]{Didi Ahmad Mulya}
\ead{didiahmadmulya1@gmail.com}

\author[brin]{Roni Muslim\corref{cp}}
\ead{roni.muslim@brin.go.id}
\cortext[cp]{Corresponding author}



\affiliation[brin]{
    organization={Research Center for Quantum Physics,  National Research and Innovation Agency (BRIN)},
    city={South Tangerang},
    postcode={15314},
    country={Indonesia}
}

\affiliation[uty]{
    organization={Department of Industrial Engineering,  University of Technology Yogyakarta},
    city={Yogyakarta},
    postcode={55285},
    country={Indonesia}
}

\begin{abstract}
We investigate the phenomena of order-disorder phase transition and the universality of the majority-rule model defined on three complex networks, namely the Barabási-Albert, Watts–Strogatz, and Erdős-Rényi networks. Assume each agent holds two possible opinions distributed randomly across the networks' nodes. Agents adopt anticonformity and independence behaviors, represented by the probability $p$, where with a probability $p$, agents adopt anticonformity or independence behavior. Based on our numerical simulation results and finite-size scaling analysis, it is found that the model undergoes a continuous phase transition for all networks, with critical points for the independence model greater than those for the anticonformity model in all three networks. We obtain critical exponents identical to the opinion dynamics model defined on a complete graph, indicating that the model exhibits the same universality class as the mean-field Ising model.
\end{abstract}

\end{frontmatter}
\section{Introduction}

One approach to understanding socio-political phenomena based on physics principles is proposing an opinion dynamics model. This approach leverages statistical physics principles to analyze and interpret the complex interactions within socio-political systems \cite{galam1982sociophysics, galam1999application,
castellano2009statistical,
galam2012socio, sen2014sociophysics,galam2017trump, schweitzer2018sociophysics}. In opinion dynamics models, the micro-level interactions between agents or individuals resemble spin interactions observed in the Ising model. Moreover, the opinion dynamics models are characterized by rules that dictate the interactions among individuals to reach a particular outcome, such as achieving a consensus state. 

\textcolor{black}{Models of opinion dynamics represent a topic of considerable interest within sociophysics. These models offer insightful perspectives on how individual opinions evolve and interact within social networks, contributing to our understanding of collective behavior in various sociological contexts. For example, the classic voter model illustrates the opinion formation process in a simplified setting. In this model, each agent possesses two potential opinions, influenced predominantly by the opinions of their immediate neighbors. Despite its simplicity, the voter model can exhibit complex behavior. Over time, the system may reach a consensus (all agents sharing the same opinion) or remain in a state of coexistence where different opinions persist indefinitely, depending on the structure of the underlying network and initial conditions \cite{holley1975ergodic}. The Axelrod model represents another sociophysical framework, elucidating the dissemination of cultural traits through local interactions. This model underscores the emergence of distinct cultural regions grounded in the principle of homophily. It demonstrates the formation and evolution of cultural domains, where the emergence of culturally similar groups leads to the development of clusters or regions characterized by shared cultural traits \cite{axelrod1997dissemination}. Two other popular models of opinion dynamics are the Sznajd model and the majority-rule model, also referred to as the Galam majority-rule model or the Galam model of opinion dynamism. These models offer distinct perspectives on how opinions form and evolve within social groups. The Sznajd Model provides an insightful representation of how pairs of like-minded individuals influence their neighbors, leading to a range of macroscopic societal states, including consensus, polarization, or fragmentation \cite{sznajd2000opinion}. The majority-rule model is constructed to replicate the dynamics through which collective opinions or decisions materialize within a group, predominantly influenced by the majority's preference \cite{galam1986majority,krapivsky2003dynamics}.}

\textcolor{black}{Various discrete and continuous opinion models have been proposed and developed to understand social aspects thoroughly. Discrete opinion models, such as the Sznajd model \cite{sznajd2000opinion}, the $q$-voter model \cite{castellano2009nonlinear}, and the Biswas-Sen model \cite{biswas2009model}, typically characterize the opinions of individuals (agents) as being represented by a limited number of distinct choices. In contrast, continuous opinion models, such as the DeGroot model \cite{degroot1974reaching} and the Deffuant–Weisbuch model \cite{deffuant2000mixing}, consider opinions as part of a continuous spectrum. Each opinion dynamics model framework possesses its advantages, depending on the social context under consideration. For instance, discrete models are more suitable for applications like voting or surveys concerning the selection of specific candidates, while continuous models are more apt for use in social psychological and sentiment analysis, where opinions are more complex and gradational.}

In addition to the nature of opinions, the network structure governing agent interactions plays a pivotal role. Individuals maintain diverse social connections in real-world social contexts, making conceptualizing opinion dynamics models within heterogeneous or complex networks more pragmatic. These include networks such as the Barabási-Albert (B-A) \cite{albert2002statistical}, Watts-Strogatz (W-S) \cite{watts1998collective}, and Erdős-Rényi (E-R) models \cite{erdHos1960evolution}, as opposed to homogeneous networks typified by regular lattices and complete graphs. Such an approach aligns more closely with the intricate and multifaceted nature of real-world social networks \cite{newman2018networks}.

From a physics perspective, several interesting physical features exist to explore, such as phase transition phenomena and universality \textcolor{black}{\cite{cardy1996scaling,crokidakis2015inflexibility, mukherjee2016disorder, oestereich2019three, calvelli2019phase, muslim2020phase, muslim2021phase, pires2022double, muslim2022opinion, schawe2022higher}}. The investigation of the phase transition phenomenon is compelling due to its capacity to elucidate shifts in social dynamics, such as consensus-status quo or agreement-disagreement, manifested in continuous and discontinuous phase transitions. 
\textcolor{black}{These features have been extensively studied in opinion dynamics models with various social scenarios and network topologies. Examples include the examination of universality in the Sznajd model on complete graphs with two distinct configurations \cite{muslim2020phase}, the analysis of the q-voter model using a pair approximation approach on multiplex networks \cite{gradowski2020pair}, the study of phase transition phenomena due to anticonformity and independence behaviors on regular lattices \cite{mulya2023destructive}, the investigation of phase transition phenomena in Barabási-Albert networks based on the Biswas–Chatterjee–Sen Model \cite{alencar2023opinion}, and the analysis of the Galam model with contrarian on complete graphs under the influence of time-varying external fields \cite{gimenez2023contrarian}.} They also help to understand macroscopically the dynamics that occur, even if the microscopic interactions among individuals can be intricate and complex.

\textcolor{black}{Another interesting opinion dynamics model is the majority-rule model, mainly when defined on complex networks. Phenomena of both continuous and discontinuous phase transitions from order to disorder have been observed in random graphs \cite{pereira2005majority, lima2008majority}, small-world networks \cite{campos2003small, luz2007majority, stone2015majority, zubillaga2022three}, and scale-free networks \cite{chen2015critical, vilela2020three, alencar2023droplet}. These studies collectively reveal a phase transition phenomenon at a specific critical point. However, to the best of our knowledge, there has been no comparative study exploring the impact of two types of noise defined within three networks simultaneously: B-A network (scale-free), W-S (small-world), and E-R (random graph) and how the universality class of the model manifests across these three networks, which is the main study in this paper. In a social context, both types of noise are correlated with social behaviors that disrupt consensus, where consensus decreases as increases the social behaviors, leading to a social phase changes at a specific threshold \cite{nyczka2013anticonformity, nowak2019homogeneous}.}

By borrowing terms from social psychology, these disruptive social behaviors are anticonformity and independence. Both behaviors are fundamentally contrary to conformity behavior, which always adopts the majority opinion. The difference is that anticonformity behavior evaluates the surrounding situation and acts oppositely, while independent behavior acts independently without any pressure from any party \textcolor{black}{\cite{willis1963two, willis1965conformity, macdonald2004expanding, nail2011development}}. The meanings of these two behaviors can minimize the complete consensus state where all agents eventually have the same unanimous opinion. \textcolor{black}{Various scenarios of opinion dynamics models with anticonformity and independence behaviors and the impact of both behaviors can be found in  Refs.~\cite{sznajd2011phase,nyczka2013anticonformity, javarone2014social, crokidakis2015inflexibility, chmiel2015phase, abramiuk2019independence, muslim2020phase, nowak2021discontinuous, civitarese2021external, muslim2021phase, muslim2022opinion,muslim4241509phase}.
For instance, in Ref.~\cite{sznajd2011phase}, the authors observed the phenomenon of continuous phase transition within the Sznajd model with independence behavior defined on a complete graph, 1D lattice, and 2D lattice. Continuous phase transition phenomena and universality classes were observed in the majority-rule model with nonconformity behavior on a complete graph, body-centered cubic (bcc) lattice, and triangular lattice \cite{crokidakis2015inflexibility}. Our recent studies also investigated the continuous phase transition phenomenon and universality classes of the Sznajd model with anticonformity and independence on a complete graph with two scenarios \cite{muslim2022opinion} and the majority-rule model with anticonformity and independence behaviors defined on a 2D lattice with square formation \cite{muslim4241509phase}. In general, the behaviors of anticonformity and independence in the mentioned models are represented by the probability parameter $p$, where agents adopt anticonformity or independence behaviors with a probability of $p$. Order-disorder phase transitions within opinion dynamics models are sometimes interpreted to understand further social phase changes, such as consensus-status quo and agreement-disagreement, as observed in several studies above.}

As mentioned earlier, understanding such social phase changes would be more intriguing if opinion dynamics models were defined on complex networks that can represent real social networks. This paper investigates destructive social behaviors that disrupt homogeneity, namely anticonformity and independence behaviors. Both behaviors can be considered as social noise which works similarly to temperature in the Ising model. These behaviors are represented by the probability $p$, where with a probability $p$, agents adopt anticonformity or independence behavior. Our preliminary results for this model on homogeneous networks such as 2D lattice, 3D lattice, and a complete graph are discussed in Ref.~\cite{mulya2023destructive}.

Our numerical simulation results show that the critical point causing the model to undergo a continuous phase transition is smaller for the model with anticonformity than for the model with independence. From a socio-political perspective, anti-conformist individuals are likelier to make the community more chaotic than independent individuals. Additionally, one can argue that the majority-minority opinions exist below the critical point, and at the critical point $p_c$, the system is in a stalemate situation or status quo. Based on finite-size scaling analysis, we obtained best-fit critical exponents that caused the collapse of all data. The critical exponents indicate that the model with anticonformity and independence on all three networks is identical and belongs to the same universality class as the opinion dynamics model defined on a complete graph and the mean-field Ising model.

\section{Model and Method}
The majority-rule model is an opinion dynamics model that assumes the majority opinion always prevails. This model makes sense when considering a group of individuals with an odd number of members interacting with each other. Suppose we assume that only the majority-rule mechanism will always win in every interaction. In that case, naturally, a homogeneous state will always occur, and the final state depends on the initial state of the system. For instance, if we examine two competing opinions (opinions A and B) within a population, the opinion with the initially larger size will ultimately prevail.

In addition to depending on the initial population size and system size, external parameters, such as the influence of an external field \textcolor{black}{(mass media)}, play a significant role in influencing the consensus state of the system \textcolor{black}{\cite{sirbu2017opinion, li2020effect, azhari2023external, muslim2024mass}}. Typically, in the absence of external influence and with a sufficiently large population size, it is observed that at an initial population fraction of $c = 0.5$, the system possesses an equal probability of evolving toward a complete consensus of either $m = +1$ (all agents have the same state or opinion up) or $m = -1$ (all agents have the same state or opinion down). The point $c = 0.5$ can be called the `critical point' or separator point between the two consensus states \cite{muslim2024mass}. Social interpretations, such as changes in consensus-status quo or vice versa, can be analyzed through changes in the order parameter (magnetization) concerning the noise parameter that disrupts regularity. This noise parameter functions similarly to temperature in the Ising model; at low temperatures, the dominant factor is the binding energy allowing all spins to align in one direction (ferromagnetic), but at high temperatures, thermal fluctuations disrupt this order, causing the material to lose its magnetization \cite{newell1953theory}.

To obtain a more relevant understanding of social phenomena, the opinion dynamics model is more realistic when analyzed on complex networks rather than homogeneous networks such as regular lattices or complete graphs. The reality is that the structure of complex networks is much more realistic than homogeneous networks, where each node has varying numbers of nearest neighbors (depicting different social relationships). As mentioned earlier, statistical physics features such as phase transitions make more sense when translated into socio-political language.

In this paper, we study the order-disorder phase transition of the majority-rule model with anticonformity and independence on three complex network models, namely B-A, W-S, and E-R networks. \textcolor{black}{The mean degree for the B-A, W-S, and E-R networks are $\langle k \rangle \approx 10, \langle k \rangle \approx 4$, and $\langle k \rangle = r (N-1)$ respectively, where $r = 0.1$, ensuring that each node has at least two directly connected neighboring nodes for all populations $N$ utilized.} We consider three agents interacting with each other, and these agents also adopt destructive social behaviors, namely anticonformity and independence. 

\begin{figure}[tb!]
    \centering
    \includegraphics[width = \linewidth]{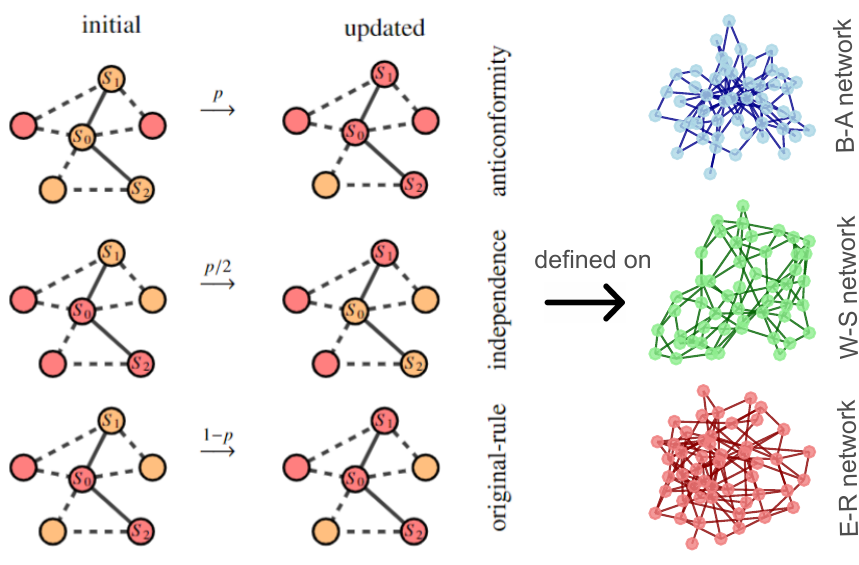}
    \caption{Illustration of the selection of three agents at random (initial and updated state): agent $S_0$ is initially chosen, followed by the random selection of its two neighboring agents, $S_1$ and $S_2$ (solid edges), who interact based on the majority-rule. The red and orange colors represent agents with opinions or states of ``up" and ``down," respectively. All agents are distributed randomly across the networks'  nodes. }
    \label{fig:skets_het}
\end{figure}

The model in this paper can be described as follows:
\begin{enumerate}
    \item Initially, the population is randomly prepared with an equal number of ``up" and ``down" opinions distributed among network nodes. 
    \item Three agents, for example, $S_0, S_1$, and $S_2$, are randomly selected to follow the majority-rule, as illustrated in Fig.~\ref{fig:skets_het}.
    \item \textcolor{black}{For the model with anticonformist agents; With a probability of $p$, agents adopt anticonformity behavior. If the three agents have the same opinion, they will change their opinions oppositely. For the model with independent agents; With a probability of $p$, agents adopt independent behavior. Then, with a probability of $1/2$, the three agents will change their opinions.}
    \item With a probability of $1-p$, the three agents will adopt the majority opinion.
\end{enumerate}

The average population opinion can be calculated using the following order parameter formulation:

\begin{equation}\label{eq:magnetization}
    m = \sum_{i =1}^{N} \dfrac{ S_i}{N},
\end{equation}
where $S_i$ represents the opinion of agent $i$ represented by the Ising number $\pm 1$. Critical exponent parameters, such as $\nu, \beta,$ and $\gamma$, are estimated using the finite-size scaling relation equation defined as follows \cite{binder1992monte}:
\begin{align}
    m & = \phi_{m}(x) N^{-\beta/\nu}, \label{eq1}\\
    \chi & = \phi_{\chi}(x) N^{\gamma/\nu}, \\
    p_c(N) - p_{c} & = c\,N^{-1/\nu}, \label{eq1b}\\
    U & = \phi_{U}(x) \label{eq2}.
\end{align}
These critical exponents apply around the critical point $p_c$ and lead to the collapse of all data $N$. The parameters $\chi$ and $U$ are susceptibility and Binder cumulant defined as follows:
\begin{align}
    \chi & = N \left(\langle m^2 \rangle - \langle m \rangle^2\right), \label{eq5}\\
    U & = 1 - \dfrac{1}{3}\dfrac{\langle m^4 \rangle}{\langle m^2 \rangle^2} \label{eq6},
\end{align}
where the critical point can be obtained from the intersection of the Binder cumulant lines $U$ and the probability $p$.

\section{Result and Discussion}

We have presented preliminary simulation results for the model in this paper, demonstrating that the model undergoes a continuous phase transition for all networks, including B-A, W-S, and E-R networks, for both the models with anticonformist and independent agents \cite{mulya2023destructive}. In that article, we did not estimate the critical points and critical exponent parameters on the three networks, which are the main focus of discussion in this paper. All results presented in this paper are outcomes of numerical simulations.

\begin{figure}[tb!]
    \centering
    \includegraphics[width = \linewidth]{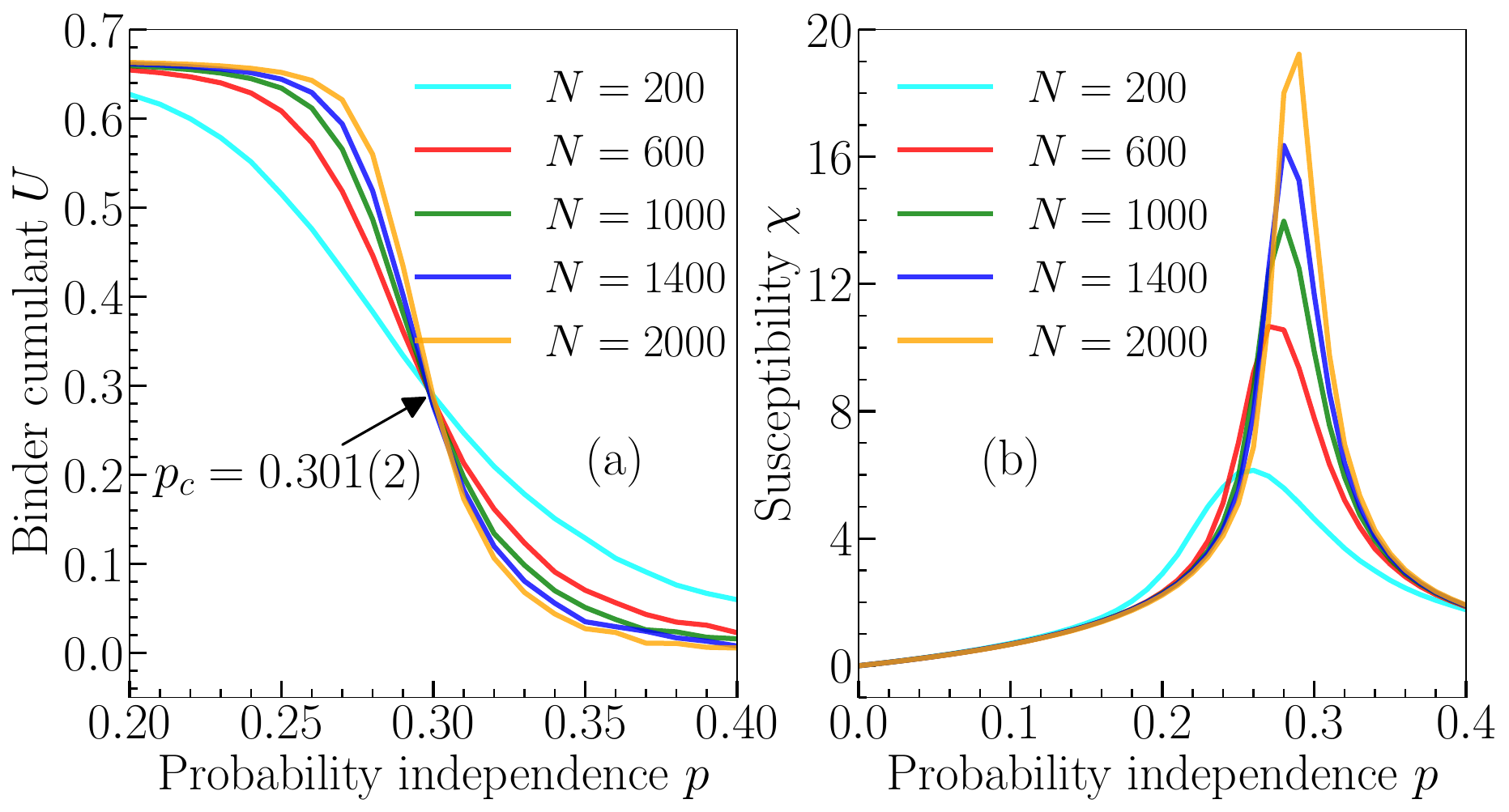}
    \caption{The numerical simulation results for the Binder cumulant $U$ [see panel (a)] and the Susceptibility $\chi$ [see panel (b)] for the model with independent agents on B-A network. The critical point of the model is determined by identifying the intersection point of the Binder cumulant $U$ versus probability $p$ curves, which is located at $p_c = 0.301(2)$. Additionally, the pseudo-critical point $p_c(N)$ for each data  $N$ is obtained from the peak of the susceptibility $\chi$.}
    \label{fig:critical_point}
\end{figure}

\begin{figure}[tb!]
    \centering
    \includegraphics[width = \linewidth]{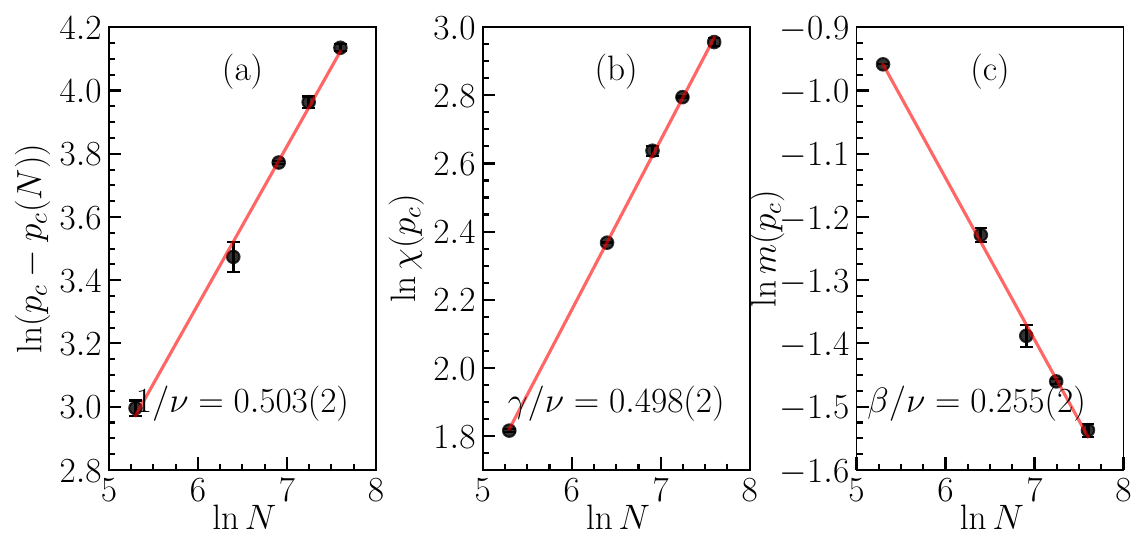}
    \caption{Log-log plot of the probability displacement $p_c-p_c(N)$ [panel (a)], susceptibility $\chi$ [panel (b)], and order parameter $m$ [panel (c)] as a function of population $N$, based on the finite-size relations in Eqs.~\eqref{eq1} - \eqref{eq1b}. All parameters are calculated at the critical point $p_c$. Critical exponents of the model are obtained from the slope of linear fitting. This plot corresponds to the model with independent agents on the B-A network.}
    \label{fig:critical}
\end{figure}

\textcolor{black}{We conducted numerical simulations with various population sizes $N$ to estimate critical points and exponents using Eqs.~\eqref{eq1} through \eqref{eq2}. The order parameter $m$ was computed by Eq.~\eqref{eq:magnetization}, while susceptibility $\chi$ and Binder cumulant $U$ were determined using Eqs.~\eqref{eq5} and \eqref{eq6} respectively. Eqs.~\eqref{eq1} through \eqref{eq1b} can be employed to obtain critical exponent values $\nu,~\beta,$ and $\gamma$. This can be achieved by utilizing the critical point value $p_c$ obtained from the intersection of lines of the Binder cumulant $U$ versus probability $p$ and calculating the pseudo-critical point $p_c(N)$ for each population size $N$ based on the maximum susceptibility $\chi$ (see Fig.~\ref{fig:critical_point}). Furthermore, the order parameter $m$ at each population's pseudo-critical point $p_c(N)$ was also obtained. To determine the ratios $1/\nu,~\gamma/\nu,$ and $\beta/\nu$, ordinary linear regression was employed to extract gradient values from the data. This approach simplifies the process of obtaining these critical exponent ratios.}
As an illustration, Fig.~\ref{fig:critical} presents the results of the finite-size scaling analysis for the model with independent agents on the B-A network. In panels (a) through (c), we obtained the following critical exponent values: \textcolor{black}{$1/\nu = 0.503(2) \rightarrow \nu = 1.987(2),$ $\gamma/\nu = 0.498(2) \rightarrow \gamma = 0.991(2),$ and $\beta/\nu = 0.255(2) \rightarrow \beta = 0.507(2)$}.

\begin{figure}[tb!]
    \centering
    \includegraphics[width = \linewidth]{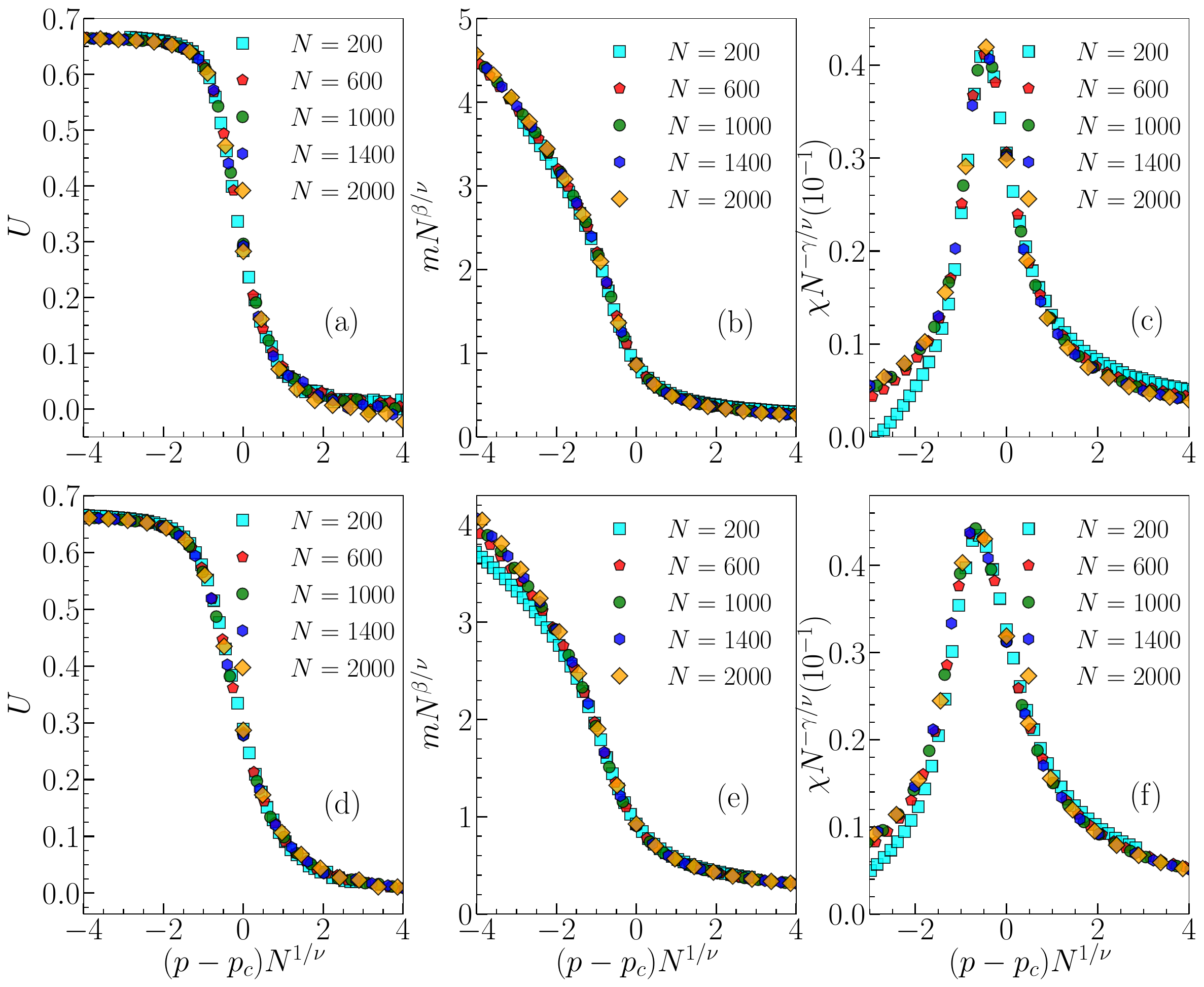}
    \caption{\textcolor{black}{Scaling plot of the order parameter ($m$), susceptibility ($\chi$), and Binder cumulant ($U$) for the models on the B-A network undergo a second-order phase transition. The critical points for the anticonformity and independence models are $p_c =  0.203(3)$ and $p_c = 0.301(2)$, respectively. The best critical exponents for the model with anticonformist agents [panels (a) to (c)] are $\nu = 2.020(2)$, $\beta = 0.509(3)$, and $\gamma = 1.016(2)$ and for the model with independent agents [panels (d) to (f)] are \textcolor{black}{$\nu = 1.987(2)$, $\beta = 0.507(2)$, and $\gamma = 0.991(2)$}. Based on these critical exponents, the models belong to the mean-field Ising universality class.}}
    \label{fig:ba_network}
\end{figure}

In general, we computed the average for each data point corresponding to $m, \chi,$ and $U$ over more than $10^6$ independent realizations to ensure the accuracy of our results.  Each data point was calculated when the simulation had reached its steady state.  Our numerical simulation results for scaling plot of $U, m,$ and $\chi$ on the B-A network are exhibited in Fig.~\ref{fig:ba_network}. Panels (a) - (c) correspond to the model with anticonformist agents, while panels (d) - (f) correspond to the model with independent agents. The critical point $p_c$ is determined from the intersection of lines of the Binder cumulant $U$ versus probability $p$, yielding \textcolor{black}{$p_c =  0.203(3)$} for the anticonformity model and \textcolor{black}{$p_c = 0.301(2)$} for the independence model. These data indicate that the models undergo a continuous phase transition at the mentioned critical points. These data also lead to the collapse of all data for different population sizes $N$ around the critical point $p_c$. A similar approach was applied to the model with anticonformist agents on the B-A network. We obtained the critical exponents as follows: \textcolor{black}{$1/\nu = 0.495(3) \rightarrow \nu = 2.020(3), \gamma/\nu = 0.503(2) \rightarrow \gamma = 1.016(2)$ and $\beta/\nu = 0.252(3) \rightarrow \beta = 0.509(3)$.}

\begin{figure}[tb!]
    \centering
    \includegraphics[width = \linewidth]{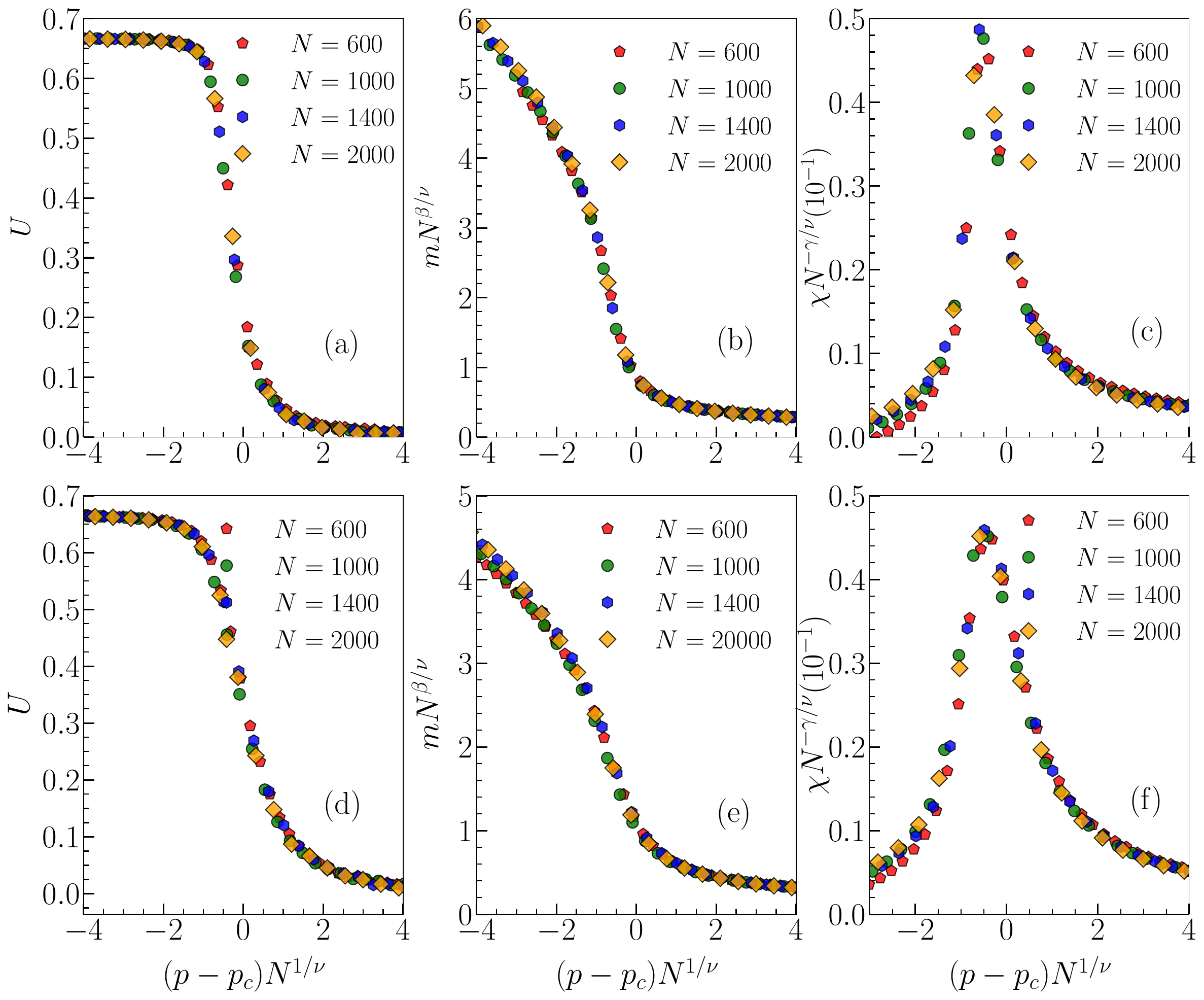}
    \caption{\textcolor{black}{Scaling plot of the order parameter ($m$), susceptibility ($\chi$), and Binder cumulant ($U$) for the models on the W-S network undergo a second-order phase transition. The critical points for the anticonformity and independence models are $p_c = 0.116(2)$ and $p_c = 0.203(2)$, respectively. The best critical exponents for the model with anticonformist agents [panels (a) to (c)] are $\nu = 2.028(5)$, $\beta = 0.501(5)$, and $\gamma = 1.001(4)$ and for the model with independent agents [panels (d) to (f)] are $\nu = 1.992(4)$, $\beta = 0.503(3)$, and $\gamma = 0.998(2)$. Based on these critical exponents, the models belong to the mean-field Ising universality class.}}
    \label{fig:ws_network}
\end{figure}

\begin{figure}[tb!]
    \centering
    \includegraphics[width = \linewidth]{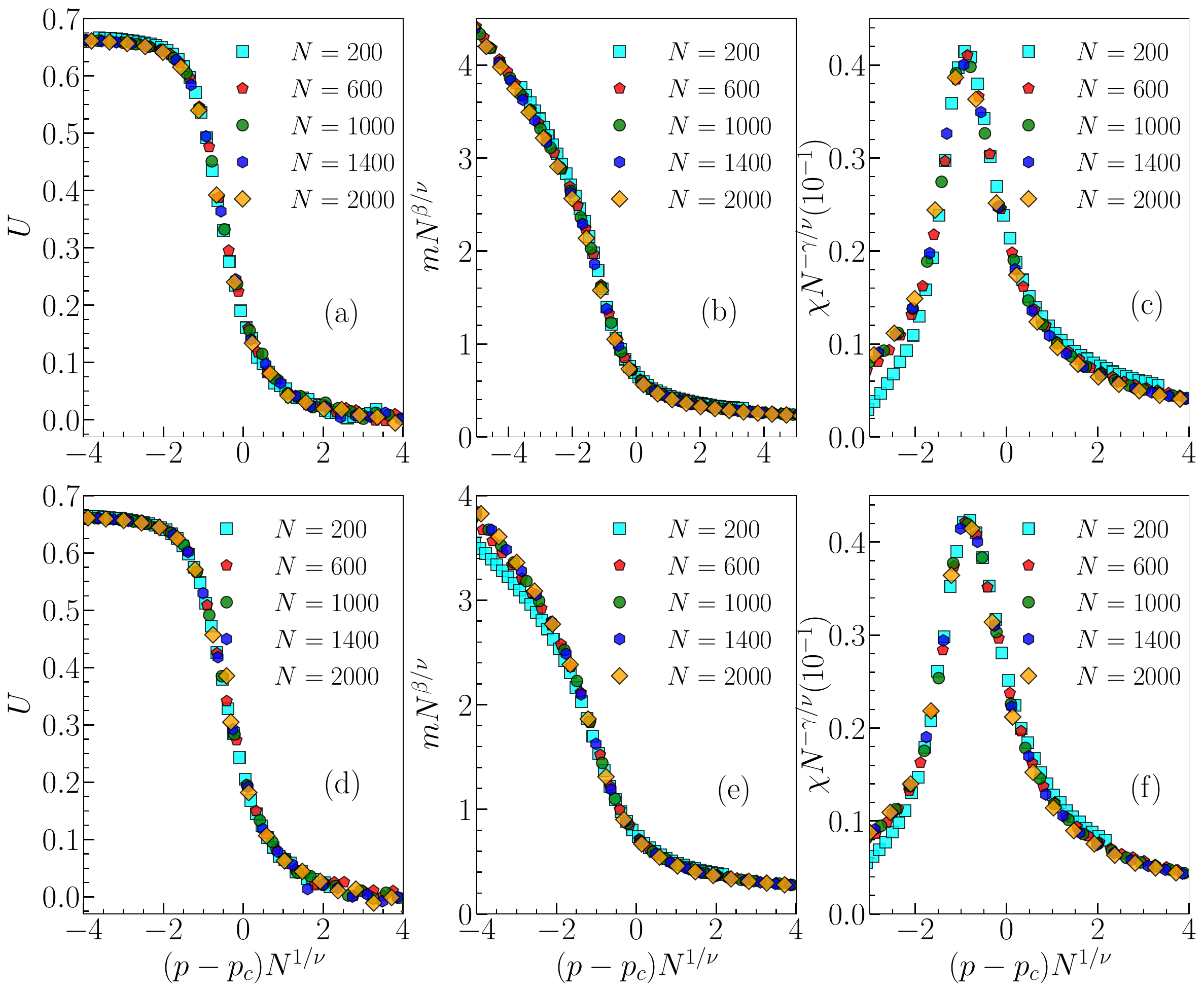}
    \caption{\textcolor{black}{Scaling plot of the order parameter ($m$), susceptibility ($\chi$), and Binder cumulant ($U$) for models on the W-S network undergo a second-order phase transition. The critical points for the anticonformity and independence models are $p_c = 0.255(3)$ and $p_c = 0.337(3)$, respectively. The best critical exponents for the model with anticonformist agents [panels (a) to (c)] agent are \textcolor{black}{$\nu = 2.012(3)$, $\beta = 0.498(4)$, and $\gamma = 0.991(2)$} and for the model with independent agents are \textcolor{black}{$\nu = 2.000(3)$, $\beta = 0.502(3)$, and $\gamma = 0.995(2)$}. Based on these critical exponents, the models belong to the mean-field Ising universality class.}}
    \label{fig:er_network}
\end{figure}

The model on W-S and E-R  networks also undergoes a continuous phase transition with critical points as follows: For the model on the W-S network with anticonformist and independent agents, the critical points are \textcolor{black}{$p_c = 0.116(2)$ and $p_c = 0.203(2)$}, respectively. For the model on the E-R network with anticonformist and independent agents, the critical points are \textcolor{black}{$p_c = 0.255(3)$ and $p_c = 0.337(3)$}, respectively. Interestingly, based on the finite-size scaling analysis, the models on the W-S and E-R networks exhibit similar critical exponents to the model on the B-A network. In detail, the critical exponents for the model with anticonformist agents on the W-S network are \textcolor{black}{$1/\nu = 0.493(5) \rightarrow \nu = 2.028(5)$, $\gamma/\nu = 0.494(4) \rightarrow \gamma = 1.001(4)$, and $\beta/\nu = 0.247(5) \rightarrow \beta = 0.501(5)$} (see panels (a) - (c) of Fig.~\ref{fig:ws_network} for the scaling plot). Furthermore, the critical exponents for the model with independent agents on the W-S network are \textcolor{black}{$1/\nu = 0.502(4) \rightarrow \nu = 1.992(4)$, $\gamma/\nu = 0.501(2) \rightarrow \gamma = 0.998(2)$, and $\beta/\nu = 0.253(3) \rightarrow \beta = 0.503(3)$ }(see panels (d) - (f) of Fig.~\ref{fig:ws_network} for the scaling plot). For the models on the E-R network, we obtain critical exponents for the model with anticonformist agents as follows: \textcolor{black}{$1/\nu = 0.497(3) \rightarrow \nu = 2.012(3)$, $\gamma/\nu = 0.493(2) \rightarrow \gamma = 0.991(2)$, and $\beta/\nu = 0.248(4) \rightarrow \beta = 0.498(4)$} (see panels (a) - (c) of Fig.~\ref{fig:er_network} for the scaling plot). Meanwhile, for the model with independent agents on the E-R network, its critical exponents are \textcolor{black}{$1/\nu = 0.498(3) \rightarrow \nu = 2.000(3)$, $\gamma/\nu = 0.496(2) \rightarrow \gamma = 0.995(2)$, and $\beta/\nu = 0.251(3) \rightarrow \beta = 0.502(3)$} (see panels (d) - (f) of Fig.~\ref{fig:er_network} for the scaling plot).

Notice that the exponents $\beta = 1/2$ and $\gamma = 1$ are typical Ising mean-field exponents, which do not apply to $\nu$. This discrepancy is associated with a superior critical dimension $d_c = 4$, so that we obtain $\nu = d_c \nu'= 2$ where $\nu' = 1/2$ is an effective exponent. Based on these critical exponents, we suggest that the model falls within the same universality class as the opinion dynamics model defined on a complete graph, as demonstrated in previous studies \cite{muslim2020phase,muslim2021phase,muslim4241509phase,muslim2022opinion, mulya2023destructive}, and the mean-field Ising model \cite{stanley1971phase}. These results imply that, despite the absence of complete connectivity among agents as in a complete graph, the model exhibits universal parameters akin to opinion dynamics models where agents are fully interconnected. 

\section{Final Remark}

We conducted an analysis of the order-disorder phase transition and the universality class of the majority-rule model implemented on heterogeneous networks, including the Barabasi-Albert, Watts–Strogatz, and Erdos-Renyi networks, wherein each node maintains connections to at least two other nodes. Each agent possesses two possible opinions, namely $\pm 1$, randomly distributed across all network nodes. Additionally, each agent adheres to the principle of conformity, characterized by behavior aligned with the majority opinion. Notably, agents also exhibit anticonformity behavior (anticonformist agents) and independence behavior (independent agents). Within this model, anticonformist agents adopt the minority opinion, adjusting their opinion or state to align with the minority viewpoint, while independent agents act autonomously, unaffected by the influence of other agents.

In a given population, three agents are selected randomly, and with a probability $p$, agents exhibit anticonformity and independence behaviors. For independent agents, the three agents change their opinions without group influence. In other situations, when all three agents hold identical opinions, they collectively alter their opinions, characteristic of anticonformist agents. Conversely, with a probability of $1-p$, the agents conform by adopting the majority opinion.

In our investigation of the phase transition, we analyzed to estimate critical exponent parameters using finite-size scaling relations to ascertain the universality class associated with the model. It was observed that the critical point for the anticonformity model is smaller in magnitude compared to that of the independence model across all network configurations (see Table~\ref{Table1}). Remarkably, for both anticonformity and independence models applied to all network topologies, the critical exponents exhibited striking similarities with those found in the opinion dynamics model defined on a complete graph and the mean-field Ising model. These results suggest that the models fall within the same universality class as the opinion dynamics model defined on a complete graph and the mean-field Ising model.

\begin{table}[ht]
\caption{Critical point and critical exponents of the model on B-A, W-S, and E-R networks}
\centering
\small
\begin{tabular}{@{}lccccc@{}}
\toprule
\multirow{2}{*}{Feature} & \multirow{2}{*}{Network} & \multirow{2}{*}{$p_c$} & \multicolumn{3}{c}{Critical exponents} \\ \cmidrule(l){4-6} 
                        & & & $\nu$ & $\beta$ & $\gamma$\\ \midrule
\multirow{3}{*}{Antic.} & B-A  & $0.203(3)$    & $2.020(2)$
  & $0.509(3)$   & $1.016(2)$   \\
                                     & W-S  & $0.116(2)$    & $2.028(5)$  & $0.501(5)$   & $1.001(4)$   \\
                                     & E-R  & $0.255(3)$    & $2.012(3)$  & $0.498(4)$   & $0.991(2)$   \\ \midrule
\multirow{3}{*}{Indep.}   & B-A  & $0.301(2)$    & $1.987(2)$  & $0.507(2)$   & $0.991(2)$   \\
                                     & W-S  & $0.203(2)$    & $1.992(4)$  & $0.503(3)$   & $0.998(2)$   \\
                                     & E-R  & $0.337(3)$    & $2.000(3)$  & $0.502(3)$   & $0.995(2)$   \\ \bottomrule
\end{tabular}\label{Table1}
\end{table}

\section*{Author Contributions:}
Both authors have contributed equally.

\section*{Declaration of Interests}
The authors affirm that they have no known competing financial interests or personal relationships that could have influenced the work reported in this paper.

\section*{Acknowledgments}
The authors express their gratitude to the BRIN Research Center for Quantum Physics for providing the mini HPC (Quantum Simulation Computer) for conducting numerical simulations. \textbf{D.~A.~Mulya} acknowledges the support received from the Research Assistant program of BRIN talent management through Decree Number 60/II/HK/2023.


\end{document}